\documentclass[twocolumn,showpacs,floatfix,amsmath,amssymb,prb,superscriptaddress]{revtex4-1}

\usepackage{graphicx}% Includeure files
\makeatletter
\graphicspath{./figuras}

\begin{document} 
\newcommand{\be}{\begin{equation}}
\newcommand{\ee}{  \end{equation}}
\newcommand{\ba}{\begin{eqnarray}}
\newcommand{\ea}{  \end{eqnarray}}
\newcommand{\ve}{\varepsilon}

\title{Pumping charge with ac magnetic fluxes and  the dynamical breakdown of Onsager symmetry}

\author{Mar\'{\i}a Florencia Ludovico} \affiliation{Departamento de F\a'{i}sica,
  FCEyN and IFIBA, Universidad de Buenos Aires, Pabell\'on 1, Ciudad
  Universitaria, 1428 Buenos Aires, Argentina}

\author{Liliana Arrachea} \affiliation{Departamento de F\a'{i}sica,
  FCEyN and IFIBA, Universidad de Buenos Aires, Pabell\'on 1, Ciudad
  Universitaria, 1428 Buenos Aires, Argentina}

\date{\today}

%%%%%%%%%%%%%%%%%%%%%%%%%%%%%%%%%%%%%%%%%%%%%%%%%%%%%%%%%%%%%%%%%%%%
\begin{abstract}
We study the transport properties of  setups with one and two mesoscopic rings threaded by
ac magnetic fluxes of the form $\Phi(t)=\Phi^{\rm dc} + \Phi^{\rm ac} \cos(\Omega_0 t + \delta)$ and
connected to two different particle reservoirs. We analyze the conditions to generate a pumped dc current in the adiabatic regime.
We also study the
 symmetry properties of the induced dc current  as a function of the  static
component of the flux, $\Phi^{\rm dc}$, with and without a dc bias voltage applied at the reservoirs. We analyze, in
particular, the validity of the Onsager-Casimir relations for different configurations of the setups.
\end{abstract}
%%%%%%%%%%%%%%%%%%%%%%%%%%%%%%%%%%%%%%%%%%%%%%%%%%%%%%%%%%%%%%%%%%%%%


\maketitle

%%%%%%%%%%%%%%%%%%%%%%%%%%%%%%%%%%%%%%%%%%%%%%%%%%%%%%%%%%%%%%%%%%%%%%%%
\section{Introduction}
A  mesoscopic ring threaded by a magnetic flux is one of the paradigmatic systems to discuss the
 fundamentals of quantum electronic transport. Several experiments \cite{1, 2, 3} and
 theoretical works \cite{4, 5} devoted to the investigation of the Aharonov-Bohm effect and related phenomena originated by static magnetic
 fluxes, constitute milestones in this area of Physics. 

In the case of a time-dependent magnetic flux, a very interesting  example corresponds to that where the flux increases linearly in time. In this case
a   constant electric field is induced along the circumference of the ring, generating a time-dependent current.
This problem was introduced by B\"uttiker, Imry and Landauer in the early times of the theory of quantum transport \cite{6} and triggered very insightful discussions 
on the role of  inelastic scattering as a necessary ingredient 
 to generate a dc component in the induced current along the ring.  \cite{7,8,9,10,11,12,13,14,15,16,17} The discussion of these effects, along with the effect of 
 disorder, and the comparison of the current generated in this setup with the current generated by a dc bias voltage were  the subject of further investigations. 
 \cite{18,19,19a,baruch} While these studies have been illuminating from the conceptual point of view, the experimental implementation of a magnetic flux with a linear
  dependence in time is not very practical. Instead, fluctuating magnetic fluxes are much more
usual in the laboratories. 

A magnetic flux with an harmonic dependence on time, characterized by a frequency $\Omega_0$, threading a ring connected to a single wire  was considered by 
B\"uttiker  in Ref. ~\onlinecite{9}. That work was developed in the framework  of the discussion of the combined effect of the electromotive force (emf), induced by the magnetic flux, and the inelastic scattering
 in  inducing a dc current along the circumference of the ring. The coupling to  the wire was introduced to provide a concrete mechanism for inelastic scattering and
  decoherence which leads to an induced current along the ring with a finite dc-component. More recently, further details of the induced current along the ring were analyzed in the
 same setup. \cite{curring} Other mechanisms to introduce decoherence in the ring without coupling to electronic reservoirs were also considered, \cite{marq} as well as
  the effect of spin-orbit coupling. \cite{spino}    
 In the case of the ring connected to a single reservoir, a  time-dependent current is induced through the contact between the ring and the reservoir as a consequence
 of the driving. Such current has a zero dc-component. However, when additional reservoirs are coupled to the ring as in the sketch of Fig. 1a,  the driven ring may 
 behave as a quantum pump that induces a current with a net dc component between the reservoirs. The aim of the present work is, precisely, to analyze the behavior of 
such a current.

 In the last years,  quantum pumps have received a lot of attention from both experimental
and theoretical communities. \cite{pumps,brouw,moska,lilipump,lilisim,nonsim,mobu,foa} The basic idea  of these setups is the generation of a dc current in the absence of an explicit dc bias. Most
of the studied devices are mesoscopic structures locally driven by ac gate voltages. The key to induce a dc current under
these conditions is the breaking of time-inversion and spacial-inversion symmetries. \cite{lilisim,20,21,22} However, depending on the
mechanism employed, different behaviors are expected for the induced dc current, as a function of $\Omega_0$.
In particular, the so called adiabatic regime, characterized by a dc current with a linear dependence in $\Omega_0$, is achieved when
pumping is induced by a setup with two time-dependent parameters, \cite{brouw} which are usually two ac potentials applied at different places of
the structure and oscillating with a phase-lag.  Pumping mechanisms have been studied in rings threaded by a static magnetic
flux and driven by applying a local ac gate voltage at some point of the circumference. \cite{nonsim,mobu,lilisim,foa} Adiabatic pumping
was proposed to be generated by two rings threaded by harmonic magnetic fluxes oscillating with a phase-lag and connected by
a tunneling contact. \cite{doub} There is also a very recent proposal of generating pumping with magnetic fluxes in Cooper Pair boxes. \cite{sim}
In the present work we study configurations containing one and two rings connected to two different particle reservoirs and threaded by magnetic fluxes with dc and ac components. The ac components of the fluxes oscillate with a frequency $\Omega_0$.
This system behaves as a quantum pump, which induces a 
dc current between the two reservoirs. 
Our first goal is to identify under which conditions an adiabatic regime is expected in this setup. We extend our analysis to study the transport properties
when an additional small bias voltage $V$ is applied at the reservoirs. We also discuss conditions and parameters relevant for the observation of these regimes.

An important related issue is the behavior of the induced dc current as a function of the static component of the flux, $\Phi^{\rm dc}$ and the corresponding
analysis of the validity of Onsager-Casimir relations in this setup. The latter result as a consequence of microreversibility and enforce the linear stationary
 conductance of a two-terminal system to be an even function of $\Phi^{\rm dc}$. 
 For large voltages $V$, beyond the linear 
response regime, there is no reason to expect that symmetry in the induced current and its breakdown has been suggested to have interesting consequences on
thermoelectric effects. \cite{ben} Several recent  works have been devoted to the study of mechanisms for breaking 
Onsager symmetry in the non-linear conductance theoretically \cite{viol-casimir-theor} as well as in several  experimental settings.\cite{viol-casimir-exp, ang} 
In most of these cases, the source for the asymmetric behavior of the current as a function of $\Phi^{\rm dc}$ was identified to be the effective voltage profile induced along the biased structure
as a consequence of the Coulomb interaction. 
 In the case of rings threaded by dc fluxes while  biased by ac voltages there are also experimental results, 
which are supported by semiclassical theoretical arguments, indicating that the Onsager-Casimir relations are in general not valid for the conductance associated 
to the rectified current. \cite{ang} In this context, the second goal of the present work is to check the validity of Onsager symmetry in setups containing rings threaded by magnetic fluxes
with dc as well as ac components. To this end, we
define appropriate conductance coefficients to characterize the dc-current and analyze the symmetry properties of these coefficients as functions of $\Phi^{\rm dc}$.

The paper is organized as follows. In section II we present the model, the theoretical treatment to evaluate the dc currents, as well as the definitions of the different transport coefficients. In section III we 
discuss the conditions to have adiabatic pumping. Section IV is devoted to analyze the symmetry properties of the pumped dc current as a function of the dc magnetic flux. This analysis is extended in
Section V, where we also consider the effect of a dc bias voltage.   In section VI we close with a summary and the conclusions.

\section{Theoretical Approach}
\subsection{Model} 
\label{model}

\begin{figure}[tb]
\includegraphics[width=0.5\textwidth]{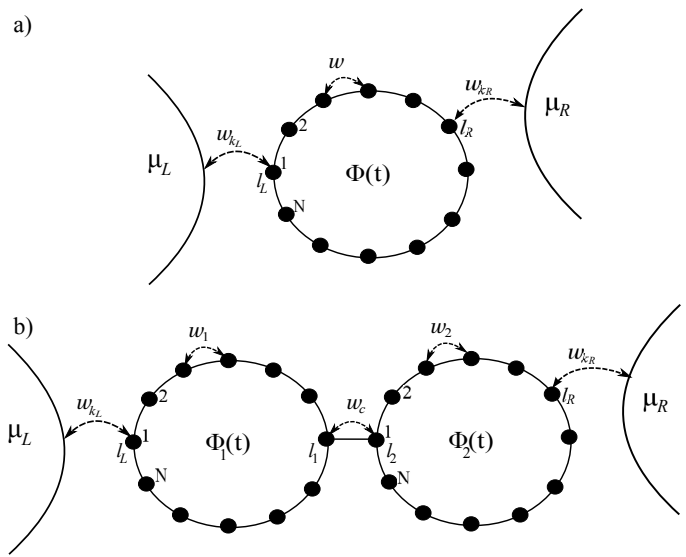}
 \caption{Sketches of the systems considered in this work. Panel a): a single metallic ring threaded by a magnetic flux $\Phi(t)$ and connected to two particle reservoirs with chemical potentials $\mu_L$ and $\mu_R$ respectively. Panel b): two connected rings threaded by magnetic fluxes $\Phi_j(t)= \Phi^{\rm dc}_j+ \Phi^{\rm ac}_j \cos(\Omega_0 t + \delta_j), \; j=1,2$.
The rings are described by tight-binding chains with $N$ sites.  The reservoirs are attached at the sites $l_L$ and $l_R$ are those sites at which the leads are attached through
tunneling contacts $w_{k_L}$ and $w_{k_R}$.}
\label{fig1}
\end{figure}

The system we consider is sketched in Fig.~\ref{fig1}. It consists in one or two single-channel rings of length $L$ threaded by harmonically time-dependent magnetic fluxes
of the form 
$\Phi_j(t)=\Phi_j^{\rm dc}+\Phi_j^{\rm ac} \cos(\Omega_0 t+ \delta_j)$.   The different rings are labeled by $j=1,2$ and are 
connected to one-dimensional left ($L$) and right ($R$) wires that play the role of reservoirs. 
The ensuing Hamiltonian is
\be
H=H_{r}(t) + \sum_{\alpha=L,R} H_{\alpha}+ H_{cont}.
\ee
The Hamiltonians for the rings correspond to tight-binding models  with lattice constant $a$ and $N$ sites each, thus $L=Na$,
\ba
H_{r} (t) & = & - \sum_{j=1}^2 w_j \sum_{l=1}^N\left[ e^{-i \phi_j(t) }c^{\dagger}_{l,j} c_{l+1,j} + H.c\right] + \nonumber \\
& &   -w_c \left[c^{\dagger}_{l_1,1} c_{l_2,2} + H.c\right] .
\ea
The phases $\phi_j(t)= 2 \pi \Phi_j(t)/L, \; j=1,2$ are  Pierls factors, which account for the magnetic fluxes threading the rings in units of the flux quantum
$\Phi_{0}=h c/e$. For simplicity, we consider spinless electrons and we impose periodic boundary conditions $N+1\equiv 1$ in each ring. The number $N$ determines the number of
discrete levels of the isolated ring within the energy range $[-2w,2w]$. Thus, $N$ also sets the typical level spacing $\Delta=4w/N$.

The last term represents the coupling between the two rings. The leads or reservoirs are represented by non-interacting Hamiltonians 
$H_{\alpha}= \sum_{k_{\alpha}} \ve_{k_{\alpha}} c^{\dagger}_{k_{ \alpha}} c_{k_{\alpha}}$. The
Hamiltonian representing the contact between the leads and the rings reads
\be
H_{cont}=\sum_{\alpha=L,R}\sum_{k_{\alpha}} w_{k_{\alpha}} \left[c^{\dagger}_{k_{\alpha}} c_{l_{\alpha}} + H. c. \right],
\ee
where $l_{\alpha}$ are the sites of the rings at which the leads are attached. The case of a single ring corresponds to $w_2=w_c=0$ in $H_r(t)$ and both
$l_{\alpha}$ lying on the first ring.
 In what follows, in order to simplify the notation, we adopt units where $e=c=\hbar=1$.  We also set to unit the lattice parameter $a$. We will restore these constants and change the units when appropriate.

\subsection{dc current and transport coefficients}\label{conductance}
In the most general case, we assume  a small voltage $V$ applied in the setup, which is represented as difference between the chemical potentials
of the  $L$ and $R$ reservoirs, respectively, $\mu_L=\mu+V$ and $\mu_R=\mu$. In order to analyze the transport properties, we follow the procedure of Refs. 
~\onlinecite{lilipump}, to which
we refer the reader for further details. The evaluation
of the dc current flowing through the contact between the reservoir $\alpha$ and the ring to which it is attached cast
\ba \label{jdc}
J^{\rm dc}_{\alpha}& = & \sum_{\beta=L,R} \sum_n \int \frac{d \omega}{2 \pi} \left[f_{\beta}(\omega)- f_{\alpha}(\omega+n \Omega_0) \right] 
\nonumber \\
& & \times \Gamma_{\alpha}(\omega + n \Omega_0) | {\cal G}_{l_{\alpha} l_{\beta}} (n, \omega )|^2 \Gamma_{\beta}(\omega),
\ea
where $\Gamma_{\alpha}(\omega)= 2 \pi \sum_{k_{\alpha}} |w_{k_{\alpha}} |^2 \delta(\omega - \varepsilon_{k_{\alpha}})$ and 
$f_{\alpha}$ is the Fermi function corresponding to the reservoir $\alpha$ in equilibrium. During our work we consider that the system is at $T=0$, thus $f_{\alpha}=\Theta (\mu_\alpha -\omega)$.

The retarded Green's function for sites $l,l^{\prime}$ of the rings (for simplicity we use a single label $l$ to identify the site and the ring) is expressed in terms of the Floquet-Fourier representation
\ba
\label{Floquet}
G^{R}_{l,l{\prime}} (t, t^{\prime}) & =  & \int  \frac{ d \omega}{ 2\pi} G^R_{l,l{\prime}} (t, \omega )e^{- i \omega (t-t^{\prime})},
\nonumber \\
G^R_{l,l{\prime}} (t, \omega ) & = & 
 \sum_n e^{-i n \Omega_0 t} {\cal G}_{l,l{\prime}} (n, \omega ).
\ea
To evaluate the latter Green function, it is convenient to express the Hamiltonian for the rings as 
\be
\hat{H}_{r}(t)= \hat{H}_0 + \hat{\cal V}(t),
\ee
where $ \hat{H}_0$ and  $\hat{\cal V}(t)=
 \sum_{n\neq 0} e^{-i n \Omega_0 t} \hat{\cal V}_n$, are matrices with elements defined by the spacial coordinates of the rings. We then
 formulate the Dyson equation as follows
\ba \label{dy}
\hat{G}^R (t, \omega ) & = &  \hat{G}^0(\omega) + \nonumber \\
& & \sum_{n \neq 0} e^{-i n \Omega_0 t} \hat{ G}^R (t, \omega + n \Omega_0 ) \hat{\cal V}_n \hat{G}^0(\omega),
\ea
where $\hat{ G}^R (t,\omega)$ denotes the matrix with elements ${ G}^R_{l,l{\prime}} (t, \omega )$, being $l, l^{\prime}$ spacial coordinates of the rings, while the stationary retarded
Green function is also expressed as a matrix 
\ba
\label{g0}
\hat{G}^0(\omega) = [\omega \hat{1} - \hat{H}_0- \hat{\Sigma}(\omega)]^{-1},
\ea 
with $ \Sigma_{l,l^{\prime}}(\omega)=
\sum_{\alpha} \delta_{l,l^{\prime}} \delta_{l,l_{\alpha}} \int (d\omega^{\prime})/(2 \pi) 
\Gamma_{\alpha}(\omega^{\prime}) /(\omega -\omega^{\prime} + i 0^{+} )$.

For a very small voltage difference $V$ and low driving frequency $\Omega_0$, the dc current flowing through the contact between the ring and the  reservoir $\alpha$ 
 can in general be expressed as
\be
\label{curr_low_freq}
J^{\rm dc}_{\alpha}   =  G_{\rm dc}^V V + G_{\rm dc}^{a}\Omega_0+   
G_{\rm dc}^{na}\Omega_0^2+ 
G_{\rm dc}^{mix} V \Omega_0 ,
\ee
which satisfies $J^{\rm dc}_L=-J^{\rm dc}_R$, as is expected from the conservation of the charge. In the above equation we define four different transport coefficients.
The coefficient $G_{\rm dc}^V$ is the usual linear dc conductance, $G_{\rm dc}^a$ is the coefficient relating the dc current with the pumping frequency in the adiabatic pumping regime,
$G_{\rm dc}^{na}$ is the non-adiabatic transport coefficient and $G_{\rm dc}^{mix}$ is a coefficient that quantifies the effect of mixing between  
the dc bias and the pumping to generate the dc current.  
The corresponding expressions are 
\ba
\label{coeff}
\begin{array}{r l l}
G^{a}_{\rm dc} & = & \sum\limits_{\beta =L,R}\sum\limits_{n}n \, \Gamma_{\beta}(\mu)\Gamma_{\alpha}(\mu) \vert
 {\cal G}_{l_{\alpha},l_{\beta}}^0(n,\mu)\vert ^2 ,\vspace{1pt}\\
G^{mix}_{\rm dc} & = & -\sum\limits_{\substack{\beta=L,R \\ \beta\neq\alpha}}\sum\limits_{n}\Gamma_{\beta}(\mu)\Gamma_{\alpha}(\mu)\left[ \right.n\,\partial_{\omega}\vert{\cal G }_{l_{\alpha},l_{\beta}}^0(n,\omega)\vert ^2\delta_{\alpha,L}\vspace{1pt}\\
& & \left.-\left. 2\text{Re}\left({{\cal G }_{l_{\alpha},l_{\beta}}^0(n,\omega)}{{\cal G }_{l_{\alpha},l_{\beta}}^{1*}(n,\omega)}\right)\right]\right|_{\omega=\mu},\vspace{1pt}\\
G^{na}_{\rm dc}& = & \sum\limits_{\beta =L,R}\sum\limits_{n}\Gamma_{\beta}(\mu)\Gamma_{\alpha}(\mu)\left[\frac{n^2}{2}\partial_{\omega}\vert{\cal G }_{l_{\alpha},l_{\beta}}^0(n,\omega)\vert ^2 \right. \vspace{1pt}\\
& & \left. +\left. 
n\,2\text{Re}\left({{\cal G }_{l_{\alpha},l_{\beta}}^0(n,\omega)}{{\cal G }_{l_{\alpha},l_{\beta}}^{1*}(n,\omega)}\right)\right]\right|_{\omega=\mu},
\end{array}
\ea 
where the matrix elements ${\cal G }_{l_{\alpha},l_{\beta}}^k(n,\omega)$ with $k=0,1$ correspond to a low-frequency expansion of the Green's function of the form
\ba
\label{expansion}
\hat{\cal G }(n,\omega)\thicksim {\hat{\cal G }}^0(n,\omega)+\Omega_{0}\,{\hat{\cal G }}^1(n,\omega).
\ea  
At this point, it is interesting to mention that, when the constants $e$ and $h$ are restored, 
  is $ G_{\rm dc}^V \propto e^2/h$ while the adiabatic coefficient is $G_{\rm dc}^a \propto e$. In the adiabatic regime, the latter is the only non-vanishing
   coefficient and is also related to the average charge through $Q_{\rm dc}= 2 \pi G_{\rm dc}^a$, with
  \begin{equation}
 Q_{\rm dc}= \tau J^{\rm dc}, 
  \end{equation}
  being $\tau= 2 \pi/\Omega_0$ the period of the oscillating flux. 
In the next sections, we will analyze the dependence of the non-vanishing pumping coefficients on the parameters and symmetries of the setup. On the other hand, we will also discuss the 
symmetry properties of the current as a function of $\Phi^{\rm dc}_j,\;j=1,2$.

\subsection{Evaluating the retarded Green's function}
In this subsection we present the different strategies that we will follow to evaluate the retarded Green's function entering the expression for the current in different 
relevant
limits. 

\subsubsection{Small ac amplitudes}
The solution of the Dyson equation (\ref{dy}) leads to the exact Green's function. For weak 
amplitudes $\phi^{\rm ac}_1$ and $\phi^{\rm ac}_2$ of the ac component of the magnetic flux and arbitrary frequency $\Omega_0$, it is possible
to solve that equation perturbatively. It is convenient to start expanding the Hamiltonian $H_r(t)$ in
powers of $\phi^{\rm ac}_1, \; \phi^{\rm ac}_2$. By keeping terms up to the first order in these parameters, we get
\ba \label{smallac}
H_r(t) & \sim &  H_0 + \sum_{j=1}^2 \sum_{l=1}^N\left[{\cal V}^{(j)}(t)  c^{\dagger}_{l,j} c_{l+1,j} + H.c \right],
\ea
with 
\ba
{\cal V}^{(j)}(t)=i w_j e^{-i \phi^{\rm dc}_j} \phi^{\rm ac}_j \cos(\Omega_0 t +\delta_j),
\ea 
and 
\ba \label{h0}
H_0 &= & -\sum_{j=1}^2 w_j \sum_{l=1}^N \left[e^{-i \phi^{\rm dc}_j} c^{\dagger}_{l,j} c_{l+1,j} + H.c \right] \nonumber \\
& &-w_c \left[c^{\dagger}_{l_1,1} c_{l_2,2}+ H.c.\right],
\ea
with
$\phi^{dc,ac}_j =2\pi \Phi^{dc,ac}_j/L, ~ j=1,2$. The corresponding perturbative solution of Eq. (\ref{dy}) reads
\ba \label{dyp}
\hat{G}^R (t, \omega ) & \sim &  \hat{G}^0(\omega) +  \sum_{n=-1, n\neq 0}^1 e^{-i n \Omega_0 t} \nonumber \\
& & \times \hat{ G}^0 ( \omega + n \Omega_0 ) \hat{\cal V}^n \hat{G}^0(\omega),
\ea
where $\hat{\cal V}^n=\hat{\cal V}^{n,(1)} + \hat{\cal V}^{n,(2)}$, with matrix elements
\be
{\cal V}^{\pm 1, (j)}_{l,l^{\prime}} = i w_j  \frac{\phi^{\rm ac}_j}{2}\left[ \delta_{l^{\prime},l+1} e^{-i \phi^{\rm dc}_j}  - \delta_{l^{\prime},l-1}  e^{ i \phi^{\rm dc}_j}  \right] , \;\;j=1,2,
\ee
being $l,l^{\prime}$ sites of the ring $j=1,2$,
 while $\hat{G}^0(\omega)$ is given by Eq. (\ref{g0}). This procedure can be systematically extended to 
consider higher order solutions in $\phi^{\rm ac}_1,\;\phi^{\rm ac}_2$.

\subsubsection{Single ring  weakly coupled to reservoirs}
For the case of a single ring, 
a possible route to calculate the retarded Green's functions, alternative to solving (\ref{dy}) 
consists in starting from the limit where the ring is completely uncoupled from the reservoirs.  
The corresponding Hamiltonian  can be recasted as
\be
H_r(t) = -w \sum_{k} \ve_k\left( \phi(t) \right) c^{\dagger}_k c_k,
\ee
where $c_k =1/\sqrt{N} \sum_{l=1}^N e^{-i k l} c_l$, with $k= 2 m \pi/N$, with $-N/2 \leq m < N/2$ and
$\ve_k\left( \phi(t) \right) = - 2 w \cos \left( k + \phi(t) \right)$.

 The exact retarded Green function for this problem is
\ba
g^R_{l,l^{\prime}}(t,t^{\prime}) & = & \frac{1}{N} \sum_k e^{-i k (l-l^{\prime})} g^R_k(t, t^{\prime}), \nonumber \\
g_k^R (t, t^{\prime}) & = & - i \Theta(t-t^{\prime}) \exp \{-i \int_{t^{\prime}}^{t} dt_1 \ve_k \left(\phi(t_1) \right) \}.
\ea
In the limit of small $\phi^{\rm ac}$, keeping terms up to the first order in this parameter, we can express
\be
g_k^R (t, t^{\prime})  =  \sum_{n=-1}^1 e^{-i n \Omega_0 t} \int \frac{ d\omega}{2 \pi} e^{-i \omega (t -t^{\prime})}
g_k(n,\omega),
\ee
being
\ba
\label{green_clean}
g_k(0,\omega) & =& g^0_k(\omega) = \frac{1}{\omega - \ve_k(\phi^{\rm dc}) + i \eta}, \nonumber \\
g_k(\pm 1, \omega) & = & \pm \frac{\phi^{\rm ac} v_k(\phi^{\rm dc})  }{2 \Omega_0} \left[ g_k^0(\omega)- g_k^0(\omega \pm \Omega_0) \right],
\ea
with
\ba \label{vk}
\ve_k(\phi^{\rm dc}) &=& -2 w \cos(k+\phi^{\rm dc}), \nonumber \\
v_k(\phi^{\rm dc}) &=& 2 w \sin( k+\phi^{\rm dc}).
\ea
The Green's function including the coupling to the leads  is the solution of the following Dyson's equation
\ba \label{dycle}
\hat{\cal G}(m,\omega) & = & \hat{ g}(m,\omega) + \sum_{n} \hat{\cal G}(m-n,\omega+n\Omega_0) \nonumber \\
 & & \times \hat{\Sigma}(\omega+n \Omega_0)
\hat{g}(n,\omega),
\ea
where the matrix  $\hat{g}(n,\omega)$ has matrix elements $g_{l,l^{\prime}}(n,\omega)$. This equation can be exactly solved or can
be used to obtain perturbative solutions in the coupling to the reservoirs and the strength of the potential profile.

\subsection{ Low-frequency expansion}
For low frequencies a solution exact up to ${\cal O}(\Omega_0)$ can be obtained by
expanding Eq. (\ref{dy}) as follows:

\ba 
\hat{G}^R(t,\omega) & \sim &\hat{G}^{(0)}(\omega) +
\hat{G}^R(t,\omega) \hat{\cal V}(t) \hat{G}^{(0)}(\omega) + \nonumber
\\ & & i \partial_{\omega} \hat{G}^R(t,\omega) \frac{d \hat{\cal V}(t)}{dt}
\hat{G}^{(0)}(\omega).
\ea

We define the frozen Green's function

\be \label{froz}
\hat{G}^f (t,\omega) = \left[ \hat{G}^{(0)}(\omega)^{-1} - \hat{\cal V}(t)
\right]^{-1},
\ee

in terms of which the exact solution of the Dyson equation at ${\cal  O}(\Omega_0)$ reads
  
\be 
\label{Green1}
\hat{G}^{(1)}(t,\omega) = \hat{G}^f (t,\omega) + i \partial_{\omega}
\hat{G}^f (t,\omega) \frac{d \hat{\cal V}(t)}{dt} \hat{G}^f(t,\omega).
\ee

As we will discuss in the next section, this approach is useful to evaluate  the transport coefficients in terms of the frozen Green's function within the adiabatic regime. This will allow us to analyze the symmetry properties of the current as a function of $\Phi^{\rm dc}_j$, and for different configurations of the attached reservoirs. 

\section{Conditions for adiabatic pumping}
Our first step is to analyze the conditions to get adiabatic pumping, which implies a non-vanishing coefficient $G^a_{\rm dc}$.
Using the Floquet-Fourier representation of  Eq.(\ref{Floquet}) in Eq. (\ref{Green1}), we can identify  the first term of the expansion (\ref{expansion}) 
\ba
{\hat{\cal G }}^0(n,\omega)=\int_0^{\tau} \frac{dt}{\tau} {\hat G}^f(t,\omega)e^{in\Omega_0t}.
\ea
Replacing this expression in the adiabatic coefficient of Eq.(\ref{coeff}), we obtain
\ba
\begin{array}{r l l}
G^{a}_{\rm dc}& = & \sum\limits_{\beta =L,R} \Gamma_{\beta}(\mu)\Gamma_{\alpha}(\mu) \times \vspace{2pt}\\
& &\frac{1}{2\pi}\int_{0}^{\tau} dt\text{Im}\left(G^f_{l_{\alpha},l_{\beta}}(t,\mu){\partial _t}G^{f*}_{l_{\alpha},l_{\beta}}(t,\mu)\right),\\
\end{array} \label{ga}
\ea
where from Eq. (\ref{froz}) we can calculate the derivative of the frozen Green function
\be
{\partial _t}\hat{G}^{f}(t,\omega)= \hat{G}^{f}(t,\omega)\partial_t \hat{\cal V}(t)\hat{G}^{f}(t,\omega).
\ee
It was shown in Ref. \onlinecite{brouw} that at least two parameters are necessary to have adiabatic pumping in ac driven systems. In what follows we
show that, in the present context, this implies at least two different  rings driven by two different magnetic fluxes. 

\subsection{Single ring}
The first question that arrises is about the possibility of implementing a driving with a magnetic flux characterized by two or more parameters in a single ring. Such a possibility would correspond to a single ring threaded by a flux containing several harmonics, of the form
$\Phi(t)=\Phi^{\rm dc}+\sum_{m=1}^M \Phi^{ac,(m)}\cos(m\Omega_0 t+\delta_m)$. In such a case, we can express
\be\label{ring}
\hat{H}_r(t)= e^{i \phi(t)}\hat{W}+ e^{-i \phi(t)}\hat{W}^{\dagger},
\ee
 with $\phi(t)=\Phi(t)/L$ and $W_{l,l^{\prime}}= -w \delta_{l^{\prime},l+1}$, with $l=1,N$
and $N+1\equiv 1$.

In the appendix A we show that the frozen Green's function in this case has the following structure
\be
\hat{G}^f(t,\omega)=\sum_{n=-\infty}^{+\infty}e^{i n \phi(t)}\hat{G}^{(n)}(\omega).
\ee
The integral in time entering (\ref{ga}) is proportional to
\be
\sum_{n,m}\int_0^{\tau} dt  \frac{d\Phi(t)}{dt} n \; e^{i(n-m) \Phi(t)} G^{(m)}_{l_{\alpha},l_{\beta}}(\mu)\left[G^{(n)}_{l_{\alpha},l_{\beta}}(\mu)\right]^*,
\ee
which vanishes for any  periodic $\Phi(t)$, implying a vanishing adiabatic coefficient $G^a_{\rm dc}$. We, therefore, conclude that it is not possible
to generate an adiabatic pumped current by applying pure harmonic magnetic fluxes in a single ring.

\subsection{Two rings}
\begin{figure}[tb]
\includegraphics[width=0.5\textwidth]{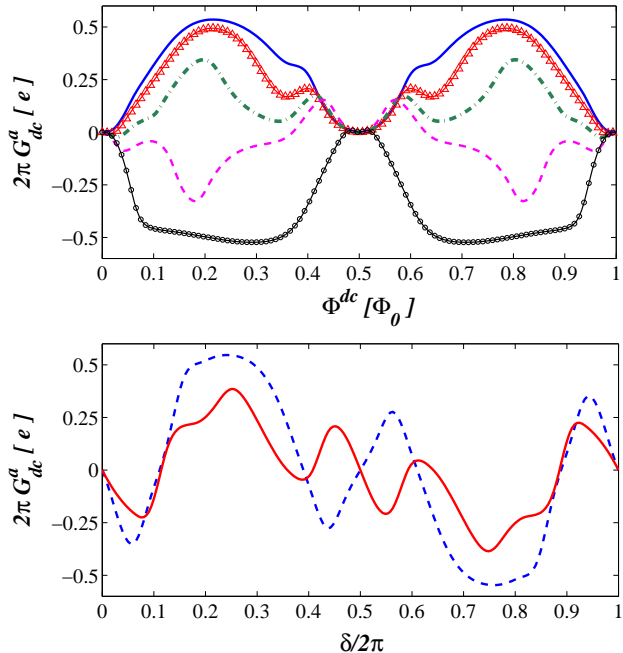}
 \caption{ Mean charge transmitted in a given period, $Q_{dc}=2\pi G^{a}_{\rm dc} $, as a function of the dc component of the magnetic flux $\Phi^{\rm dc}$, for two rings with $N=10$ sites connected at $l_1=6$ and $l_2=1$ with $w_c=1$.  The leads are symmetrically coupled to the ring with $w_L^2=0.5$ and $w_R^2=0.2$ and $N=10$. The chemical potentials are the same for the two reservoirs, $\mu_L=\mu_R=0.18$, which lies between
 two resonant levels of the ring when $\Phi^{\rm dc}=0$. The energies are in units of the hopping matrix element $w$. We consider $\Phi^{\rm dc}=\Phi^{\rm dc}_1=\Phi^{\rm dc}_2$.
  The ac flux of the left ring is fixed at the value $\Phi_1^{\rm ac}=0.32$, while $\Phi_2^{\rm ac}$ is varied and the phases of the ac fluxes are $\delta_1=0$ and $\delta_2=\pi/2$. Solid line, triangles, dot-dashed line, dashed line and circles correspond, respectively, to $\Phi_2^{\rm ac}=0.7, 0.6, 0.56, 0.47, 0.32$.
Bottom panel: $2\pi G^{a}_{\rm dc}$ as a function of the phase difference of the ac fluxes $\delta_2-\delta_1=\delta$ divided by $2\pi$. The ac components of the fluxes are $\Phi_1^{\rm ac}=0.32$ and $\Phi_2^{\rm ac}=0.8$, and the dc fluxes of the rings are equal. Dashed and solid lines correspond to $\Phi^{\rm dc}=0.22, 0.6$. All the fluxes are expressed in units of the flux quantum $\Phi_{0}=hc/e$.  }
\label{fig2}
\end{figure}

Using the frozen Green's function for small ac amplitudes of the appendix B in Eq. (\ref{coef}) leads to the following adiabatic coefficient
\be \label{2r}
G^a_{\rm dc}=\frac{\lambda^{(1)} }{2\pi} \int_0^{\tau} dt \phi^{\rm ac}_1(t) \frac{d \phi^{\rm ac}_2(t)}{dt}+
\frac{\lambda^{(2)} }{2\pi} \int_0^{\tau} dt \phi^{\rm ac}_2(t) \frac{d \phi^{\rm ac}_1(t)}{dt},
\ee
with
\ba
\lambda^{(1)}&=& \sum_{\beta=L,R} \Gamma_{\beta}(\mu) \Gamma_{\alpha}(\mu) 
\mbox{Im}\left[ \Lambda_{1}(\mu)_{l_{\alpha},l_{\beta}}  \Lambda_2(\mu)_{l_{\alpha},l_{\beta}}^* \right],
\nonumber \\
\lambda^{(2)}&=& \sum_{\beta=L,R} \Gamma_{\beta}(\mu) \Gamma_{\alpha}(\mu) 
\mbox{Im}\left[ \Lambda_2(\mu)_{l_{\alpha},l_{\beta}}  \Lambda_1(\mu)_{l_{\alpha},l_{\beta}}^* \right],
\ea
with
\be \label{lam}
\hat{\Lambda}_j(\mu)= \hat{G}^0(\mu) \hat{J}_j\hat{G}^0(\mu),
\ee
 where $\hat{J}_j$ is the current operator along the circumference of the ring $j$, defined in (\ref{cur}).
These equations  allow us to identify two conditions for a non-vanishing adiabatic current. In fact,
after performing the calculations of Eq. (\ref{2r}) explicitly, 
 it is easy to verify that
$G^a_{\rm dc} \propto \phi_1^{\rm ac} \phi_2^{\rm ac} \sin(\delta_1-\delta_2)$. Thus, the first condition  is a phase difference
for the ac fluxes,
$\delta_1 - \delta_2 \neq n \pi$.  The other condition is determined by requesting non
vanishing matrix elements $\Lambda_j(\mu)_{l_{\alpha},l_{\beta}} $, which implies  dc magnetic fluxes 
threading both rings satisfying $2 \pi \Phi_j^{\rm dc} \neq n \pi$. These two conditions are in agreement with the results of Ref. \onlinecite{doub}.
In Fig. \ref{fig2} we show that these conditions also hold for arbitrary (not necessarily small) amplitudes of the ac components. We observe that
as the amplitude of the ac components increase, $G^a_{\rm dc}$ as a function of $\delta_1 - \delta_2$ shows a much more complex structure than
the simple sinusoidal law predicted by the small amplitude result of Eq. (\ref{2r}). However, the condition of a vanishing value of this coefficient 
when the phase difference coincides with an integer value of $\pi$ is verified in all the cases. As a function of the magnetic flux, it is also
observed that for increasing ac amplitudes, there are sign changes in the behavior of the pumped current, but this current vanishes for 
$2 \pi \Phi_j^{\rm dc} = n \pi$. Further details of the behavior of $G^a_{\rm dc}$ as a function of $\Phi_j^{\rm dc}$ will be analyzed in the next section.

\section{Dependence of the pumped current on the static magnetic flux}
\subsection{Single ring}
As discussed in the previous section, it is not possible to have in this case a pumped current within the adiabatic regime. The non-adiabatic transport
coefficient $G^{na}_{\rm dc}$ is, however, non-vanishing and we now turn to study its behavior as a function of the dc magnetic flux. In particular, we are interested in
analyzing if the non-adiabatic pumped current has a defined parity as a function of the dc magnetic flux, as in the case where Onsager-Casimir relations are valid. 

In the limit of a weak coupling between the ring and the reservoirs and for small amplitudes of the ac fluxes, it is possible to find an analytical expression 
for $G^{na}_{\rm dc}$. Evaluating the dc current $J^{\rm dc}_{\alpha}$ at the lowest order in
the couplings $|w_{\alpha}|^2$, corresponds to considering $\hat{\Sigma}(\omega) \rightarrow 0$ in (\ref{dycle}), which implies
 $ \hat{\cal G}(m,\omega) \sim \hat{ g}(m,\omega) $. For small $\Omega_0$, performing 
the expansion  of Eq. (\ref{expansion}) in these Green's  functions cast
\ba
{\cal{G}}_{l_\alpha,l_\beta}^0(\pm1,\omega)=-\frac{\phi^{\rm ac}}{2N}\sum\limits_k v_k(\phi^{\rm dc})\frac{dg_k^0}{d\omega}e^{-ik(l_\alpha-l_\beta)},\nonumber\\
{\cal{G}}_{l_\alpha,l_\beta}^1(\pm1,\omega)=\mp\frac{\phi^{\rm ac}}{4N}\sum\limits_k v_k(\phi^{\rm dc})\frac{d^2g_k^0}{d{\omega}^2}e^{-ik(l_\alpha-l_\beta)}.
\ea
Replacing these expressions in the non-adiabatic coefficient (\ref{coeff}) results
\ba
\begin{array}{r l l}
\label{gna}
{G}^{na}_{\rm dc}& =&\Gamma_L(\mu) \Gamma_R(\mu) \frac{(\phi^{\rm ac})^2}{4N^2}\sum\limits_{k,k'}v_k(\phi^{\rm dc})v_{k'}(\phi^{\rm dc})\times \\
& & \left. 2 \text{Re}\left[\frac{dg_k^0}{d\omega}\frac{d^2g_{k'}^{0*}}{d\omega^2}\left(1+e^{i(k-k')(l_L-l_R)}\right)\right]\right\vert_{\omega=\mu},
\end{array}
\ea
where $l_L $ and $l_R$ are the sites of the ring where the left and right reservoirs are connected. The dependence of this coefficient on the dc magnetic flux is through the 
energies  $\ve_k(\phi^{\rm dc})$  and the currents  $v_k(\phi^{\rm dc})$  given in  Eq. (\ref{vk}). In the case of reservoirs symmetrically coupled to the ring, we have $l_R-l_L=N/2$,
which corresponds to $e^{i(k-k')(l_L-l_R)}=e^{i n \pi}$. The sums in (\ref{gna}) are, thus invariant under the change $k \rightarrow -k $ and 
$k^{\prime} \rightarrow - k^{\prime}$. Since $\ve_{-k}(-\phi^{\rm dc})=  \ve_k(\phi^{\rm dc})$ and  $v_{-k}(-\phi^{\rm dc})=  - v_k(\phi^{\rm dc})$  we can conclude that the non-adiabatic
coefficient $G^{na}_{\rm dc}$ is an even function of the dc magnetic flux $\Phi^{\rm dc}$ when the reservoirs are symmetrically connected. However, when the reservoirs are connected at arbitrary
positions the phase factor $e^{i(k-k')(l_L-l_R)}$ is no longer invariant under inversions of $k$. Then, the coefficient $G^{na}_{\rm dc}$ does not have a defined symmetry as a function of $\Phi^{\rm dc}$.

\begin{figure}[tb]
\includegraphics[width=0.5\textwidth]{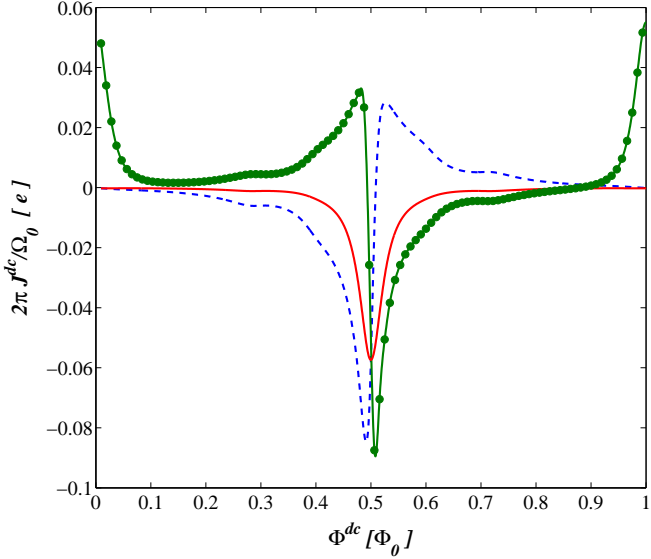}
 \caption{ The pumped current for a single ring multiplied by the period of the oscillating flux $2\pi/\Omega_0$, as a function of the dc component of the flux $\Phi^{\rm dc}$. This magnitude is related to the mean charge transmitted in a given period. The driving frequency $\Omega_0$ is resonant with the typical level spacing $\Delta$ of the ring, $\hbar \Omega_0/\Delta=1$. 
 The ac component of the magnetic flux is $\Phi^{\rm ac}=0.03$ and $\delta=0$. The fluxes are in units of $\Phi_{0}$. Solid line corresponds to wires coupled at symmetrical positions of the ring, while dashed line and circles correspond to asymmetric coupling with $l_R-l_L=4$ and $l_R-l_L=7$, respectively. Other details are the same as in Fig. 2.}
\label{fig3}
\end{figure}

In Fig. \ref{fig3} we show the behavior of the non-adiabatic pumped current as a function of $\Phi^{\rm dc}$ for a ring threaded by a flux with also an ac component $\Phi^{\rm ac}$ and reservoirs
attached at symmetric and asymmetric positions, obtained by numerically solving the Dyson equation (\ref{dy}) in the limit of a small $\phi^{\rm ac}$. In order to have a non-vanishing pumped current, it is necessary to break the spacial inversion symmetry, 
 \cite{lilisim} which in our case is accomplished by connecting the reservoirs with different tunneling amplitudes 
$w_L \neq  w_R$. These results show that the above conclusion on the behavior of the pumped current as a function of $\Phi^{\rm dc}$ is also valid for arbitrary couplings between the ring and
the reservoirs.  In fact, the pumped current is an even function of $\Phi^{\rm dc}$ 
 when the reservoirs are coupled at symmetrical positions along the ring, and it has no particular symmetry when the coupling is asymmetrical.

 Another remarkable feature that is  observed in  some cases with asymmetric coupling to the reservoirs (see for instance the plot in circles of  Fig. \ref{fig3}) is the fact that, as a function of the driving frequency $\Omega_0$, the dc current changes from
a paramagnetic-like behavior, characterized by a vanishing magnitude at $\Phi^{\rm dc}=0$ at low $\Omega_0$ to a diamagnetic-like
behavior, characterized by a sizable amplitude at $\Phi^{\rm dc}=0$. We have verified that such a change in the behavior takes place 
for a driving frequency $\hbar\Omega_0/\Delta \sim 1 $, which corresponds to resonance with the mean level spacing of the ring and it is then likely to be caused by interference effects introduced by the ac driving.

\subsection{Two rings}
We showed in the previous section that, in the case of two rings it is possible to generate a finite pumped current within the adiabatic regime. The corresponding transport coefficient is given in Eq.  (\ref{2r}) and we now focus on its symmetry properties as a function
of the dc components of the fluxes $\Phi_j^{\rm dc}$. The dependence of $G^a_{\rm dc}$ on $\Phi_j^{\rm dc}$ is enclosed in the coefficients $\lambda^{(j)}$
through the matrix elements $\Lambda_j(\mu)_{l_{\alpha},l_{\beta}}$.  This matrix is defined in Eq. (\ref{lam}). The kinetic energy 
operator of each ring, $\hat{W}_j e^{i \phi_j^{\rm dc}}+ \hat{W}_j^{\dagger} e^{-i \phi_j^{\rm dc}}$ entering the 
Green's function
$\hat{G}^0(\omega)$ is an even function under the transformation ${\cal S}_j: \phi_j^{\rm dc} \rightarrow - \phi_j^{\rm dc}, \;( l,j) \rightarrow  (-l,j) $,
corresponding to a simultaneous inversion of the flux $\phi_j^{\rm dc}$ and a spacial inversion along the circumference of the ring, while the current operator $\hat{J}_j$  is odd under such transformation. Notice, however, that the contacts between the rings as well as the contacts between the rings and
the reservoirs also enter $\hat{G}^0(\omega)$. As a consequence, ${\cal S}_j$ are symmetries of the full setup provided that the couplings between the rings and/or between rings and reservoirs do not break them. 

\begin{figure}[tb]
\includegraphics[width=0.5\textwidth]{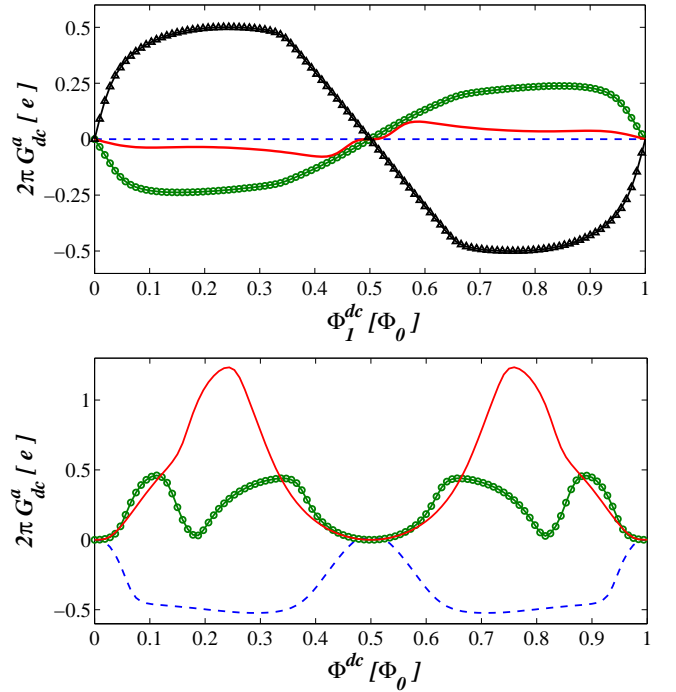}
 \caption{Mean charge transmitted in a given period, $Q_{dc}=2\pi G^a_{\rm dc}$, for two rings driven by harmonic fluxes and symmetrically connected to the reservoirs, as functions of the dc components of the fluxes (expressed in units of $\Phi_{0}$).  In both panels, the phases of the ac fluxes are $\delta_1=0$ and $\delta_2=\pi/2$.
Top panel: the dc flux of the right ring is kept fixed and the dc flux of the left ring $\Phi_1^{\rm dc}$ is varied. The ac fluxes are $\Phi_1^{\rm ac}=\Phi_2^{\rm ac}=0.32$. Dashed line, circles, solid line and triangles correspond, respectively, to $\Phi_2^{\rm dc}=0, 0.1, 0.47, 0.85$.
Bottom panel: the two dc fluxes are simultaneously changed $\Phi_1^{\rm dc}=\Phi_2^{\rm dc}=\Phi^{\rm dc}$. Solid line, circles, and dashed line correspond to $\Phi_1^{\rm ac}=\Phi_2^{\rm ac}=0.6, 0.47, 0.32$. Other details are the same as in Fig.\ref{fig2}. }
\label{fig4}
\end{figure}

\begin{figure}[tb]
\includegraphics[width=0.5\textwidth]{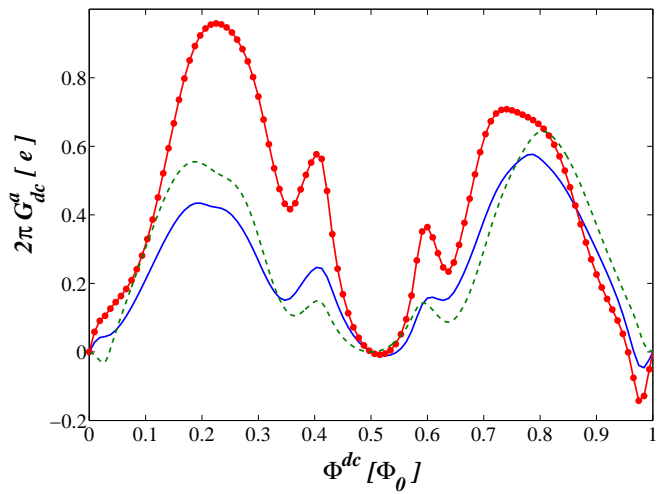}
 \caption{$Q_{dc}=2\pi G^a_{\rm dc}$ for two rings driven by harmonic fluxes asymmetrically connected to the reservoirs, as functions of the dc components of the fluxes. The two dc fluxes are simultaneously changed $\Phi_1^{\rm dc}=\Phi_2^{\rm dc}=\Phi^{\rm dc}$. The ac components of the fluxes are $\Phi_1^{\rm ac}=0.32$ and $\Phi_2^{\rm ac}=0.6$, and the phases are $\delta_1=0$ and $\delta_2=\pi/2$. The fluxes are in units of $\Phi_{0}$. Circles corresponds to wires coupled at $l_L=10$ and $l_R=2$, dashed line corresponds to $l_L=5$ and $l_R=9$, while solid line corresponds to $l_L=1$ and $l_R=2$.
 Other parameters parameters are the same as in Fig.\ref{fig2}.}
\label{fig5}
\end{figure}

Interestingly, for symmetrically connected rings and reservoirs, the transport coefficient $G^a_{\rm dc}$ and the corresponding adiabatic current
are odd functions of each of the dc fluxes,  considered independently one another, as illustrated in the top panel of Fig. \ref{fig4}, while they are even functions of the dc magnetic flux 
when it is simultaneously varied in the two rings, as shown in the bottom panel of Fig. \ref{fig4}. For arbitrary couplings between the rings and reservoirs,
the pumped current does not have any particular symmetry as a function of the dc fluxes, meaning that Onsager symmetry 
is not expected to be observed in this general case. This is illustrated in Fig. \ref{fig5} where the adiabatic transport coefficient is shown
for the case of two rings asymmetrically coupled to reservoirs.  

\section{dc current with ac fluxes and bias voltage}
We complete the analysis of the symmetry properties of the dc current as a function of the dc magnetic flux by adding the effect of
a bias voltage applied as a chemical potential difference at the two reservoirs. 

\subsection{Single ring}
In this case, we have shown in Section III that the adiabatic coefficient is $G^a_{\rm dc}=0$. A similar analysis cast $G^{mix}_{\rm dc}=0$, which means that
for small $V$ and $\Omega_0$, the voltage and the pumping contribute independently to the dc current. It is well known that
the linear stationary conductance obeys Onsager-Casimir relations, irrespectively the details of the contacts to the reservoir. Thus, from the analysis of the previous section we conclude that the full dc current of the driven ring presents Onsager-Casimir symmetry
only for the case of reservoirs that are symmetrically connected. 

\begin{figure}[h!]%[tb]
\includegraphics[width=0.5\textwidth]{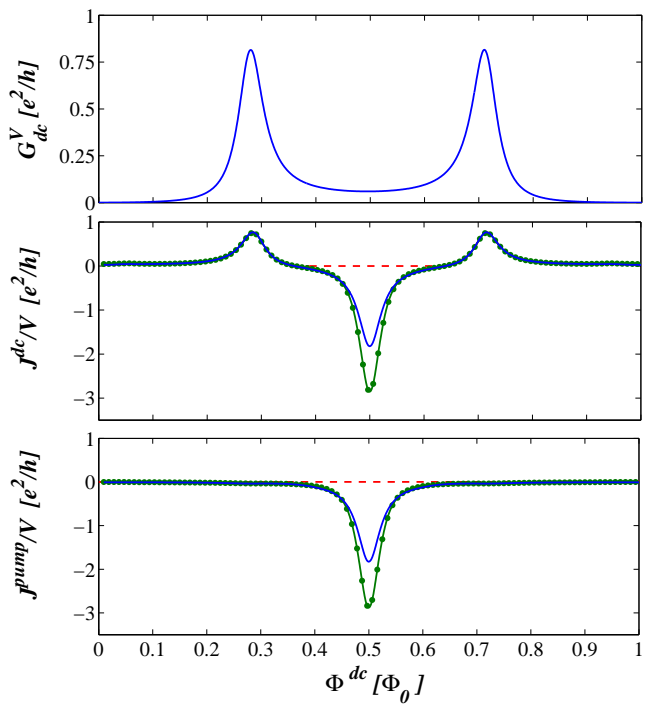}
 \caption{Upper panel  $G^{V}_{\rm dc}$ in units of $e^2/h$ as a function of the dc component of the magnetic flux $\Phi^{\rm dc}$, for a single ring
with $N=20$ sites. The chemical potentials are $\mu_{R}=\mu+eV$ and $\mu_{L}=\mu$, with $\mu=0.18$ and $eV/\Delta=0.005$.  The reservoirs are  symmetrically attached to the ring, and coupling parameters are the same as in Fig. \ref{fig3}.
In the middle and lower panels dashed line, circles, solid line correspond, respectively, to $ \hbar\Omega_0/\Delta=0.4,0.7,1$. Lower panel: pumped dc current divided by the voltage $V$ for $\mu_L=\mu_R=\mu$. The dc current resulting from the combined effect of ac driving and dc bias voltage is shown in the middle panel, divided by the bias voltage. Other details are the same as in Fig. \ref{fig3}. }
\label{fig6}
\end{figure}

\begin{figure}[h!]%[tb]
\includegraphics[width=0.5\textwidth]{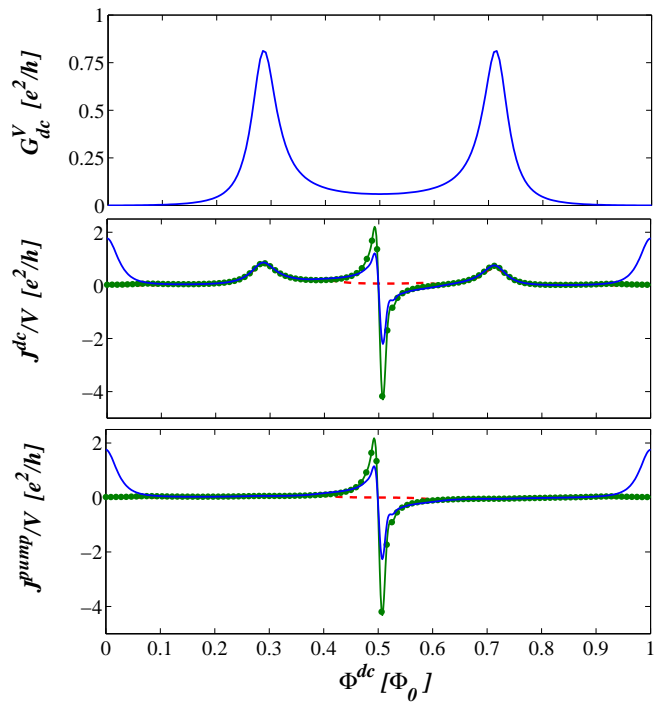}
 \caption{The same as Fig. \ref{fig6} for reservoirs  asymmetrically attached at $l_L=1$ and $l_R=10$.}
\label{fig7}
\end{figure}

In  Fig. \ref{fig6}, we show the behavior of  the total dc current
 $J^{\rm dc}= G^V_{\rm dc} V + J^{pump}$,
as a function of the dc magnetic flux, for a ring with
symmetrically connected reservoirs under the combined effect of a small bias voltage and a magnetic flux oscillating with a small amplitude $\Phi^{\rm ac}$ and several frequencies $\Omega_0$ beyond the adiabatic regime.  For the lowest frequencies $ J^{pump} \sim G^{na}_{\rm dc} \Omega_0^2$, as discussed in section III.
In the upper panel we show the linear dc conductance $G^V_{\rm dc}$, which is an even function of $\Phi^{\rm dc}$, in the middle panel the total current $J^{\rm dc}/V$.
In the latter case we divide the current by the voltage in order to show a quantity having the same units as the usual linear conductance $G^V_{\rm dc}$ shown in the top panel. 
For the symmetric connection, the pumped current $J^{pump}$ for vanishing bias voltage $V=0$ is an even function of $\Phi^{\rm dc}$ for all the frequencies considered  (see the bottom panel of  Fig. \ref{fig6}). Therefore, the total current $J^{\rm dc}$ is also an even function of $\Phi^{\rm dc}$.
 The corresponding behavior for asymmetric connections of the reservoirs is shown in Fig. \ref{fig7}. In this case, although the
linear conductance $G^V_{\rm dc}$ shown in the top panel is an even function of $\Phi^{\rm dc}$, the pumped current, shown in the bottom panel,
does not have a well defined symmetry. Thus, the total dc current shown in the middle panel of Fig. \ref{fig7} does not have a well defined symmetry as a function of $\Phi^{\rm dc}$ either.  

Notice that the quantity $J^{\rm dc}/V$ shown in the middle panels of Figs. \ref{fig6} and \ref{fig7}  can be significantly larger than the conductance quantum $e^2/h$ expected for a single channel system,
like the ones shown in the top panels where the maxima correspond to $\sim 0.8 e^2/h$. This behavior has been also discussed in the context of purely ac driven systems  (see Refs. \onlinecite{19a,sau}).
In the present case it  is a consequence of the fact that the dc current is not only induced by the bias voltage but 
also contains a non-vanishing component due to the ac driving.

\subsection{Two rings}
\begin{figure}[tb]
\includegraphics[width=0.5\textwidth]{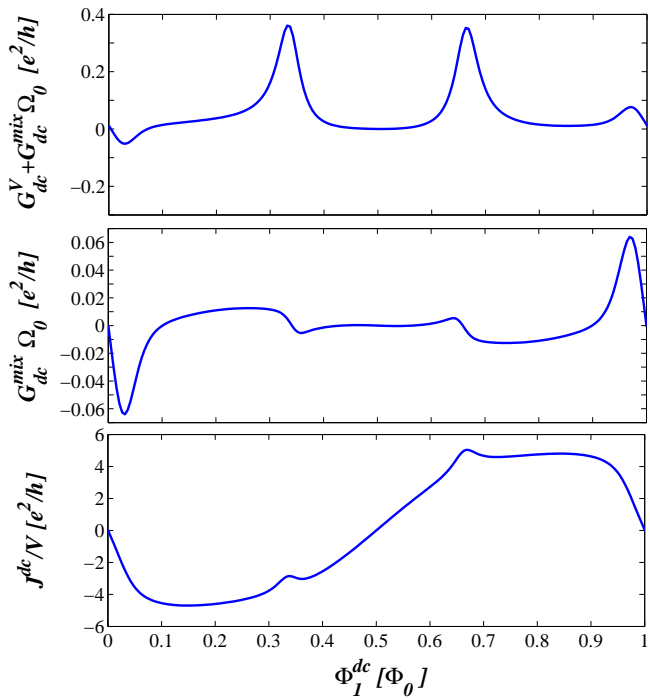}
 \caption{Upper panel: The dc current resulting from the dc bias voltage divided by the voltage $V$, $G^V_{\rm dc}+G^{mix}_{\rm dc}\Omega_{0}$, for two rings symmetrically connected. The dc component of the magnetic flux of the left ring $\Phi^{\rm dc}_1$ is varied and the one threading the second ring is kept fixed $\Phi^{\rm dc}_2=0.1$.  Middle panel: The transport coefficient $G^{mix}$ multiplied by the frequency $\Omega_{0}$. Lower panel: the total dc current resulting from the ac driving combined with the effect of the dc bias voltage divided by the voltage $V$, when $\hbar\Omega_{0}/\Delta=0.05$. In this case, since we consider identical rings, $\Delta$ corresponds to the typical level spacing of one of the rings. The dc bias voltage is $eV/\Delta=0.0025$, and the ac components of the fluxes are $\Phi_1^{\rm ac}=\Phi_2^{\rm ac}=0.32$. The phases of the ac fluxes are $\delta_1=0$ and $\delta_2=\pi/2$. Other parameters are the same as in Fig. \ref{fig2}.  }
\label{fig8}
\end{figure}

\begin{figure}[tb]
\includegraphics[width=0.5\textwidth]{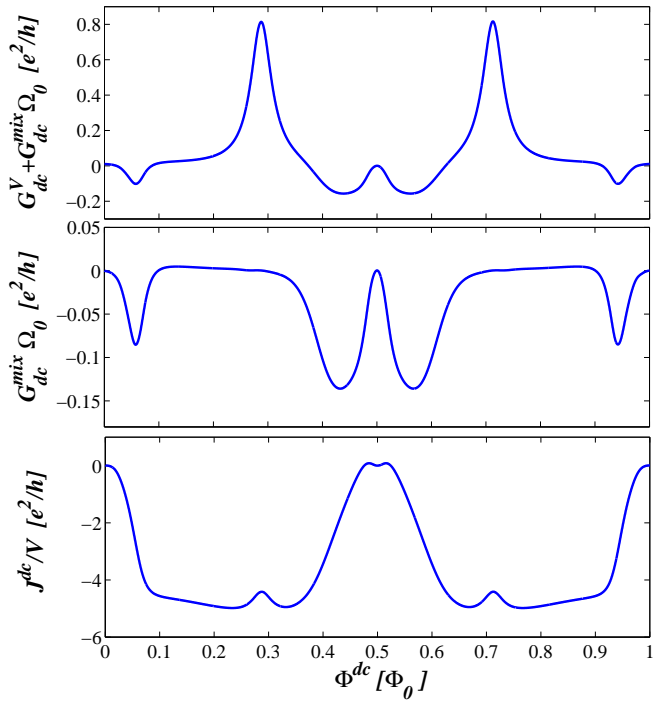}
 \caption{The same as Fig. \ref{fig8} for simultaneous variations of the dc magnetic fluxes of the two rings $\Phi^{\rm dc}_1=\Phi^{\rm dc}_2=\Phi^{\rm dc}$.  }
\label{fig9}
\end{figure}

In the case of two rings under the combined effect of ac driving and dc bias, the full dc current for small $V$ and $\Omega_0$ can be expressed as
$J^{\rm dc}=G^V_{\rm dc} V + \Omega_0 G^{a}_{\rm dc}+ \Omega_0 V G^{mix}_{\rm dc}$. In terms of the frozen Green's function the mixed transport coefficient can be writen as
\ba
\begin{array}{r l l}
G^{mix}_{\rm dc}& = &- \sum\limits_{\substack{\beta=L,R \\ \beta\neq\alpha}}\Gamma_{\beta}(\mu)\Gamma_{\alpha}(\mu)\vspace{2pt}\\
& & \int_{0}^{\tau}\frac{dt}{2\pi}\left.{{\partial}_\omega}\text{Im}\left(G^f_{l_{\alpha},l_{\beta}}(t,\omega){\partial _t}G^{f*}_{l_{\alpha},l_{\beta}}(t,\omega)\right)\right|_{\omega=\mu}.
\end{array}
\ea
In the limit of small amplitudes $\phi^{\rm ac}_j$ a similar treatment and analysis to the one performed in section III leads us to the conclusion that
this transport coefficient has the same symmetry properties as a function of $\Phi^{\rm dc}_j$ as $G^a_{\rm dc}$. Namely, for a symmetrical configuration of rings and reservoirs, this coefficient is
odd under inversion of one of the fluxes and even under the simultaneous inversion of the two fluxes. On the other hand, $G^V_{\rm dc}$ is in this case an even function under the inversion of any of the fluxes $\Phi_j^{\rm dc}$, and also under the simultaneous inversion of the two fluxes. Therefore, in the presence of a bias voltage and in a setup symmetrically connected, we expect a total dc current with no particular symmetry when a single flux is inverted, while we expect a dc current which is even under the simultaneous inversion of the two fluxes. This is illustrated in Fig. \ref{fig8}  where we consider ac fluxes with the same amplitudes driving both rings. We fix the dc component of the flux threading the right ring and analyze the dc current as the flux threading the left ring changes. We show in the top panel of the figure the dc current resulting from the bias voltage divided by the voltage $V$, $G^V_{\rm dc}+G^{mix}_{\rm dc}\Omega_{0}$. The middle panel shows the contribution of the mixed coefficient $G^{mix}_{\rm dc}\Omega_{0}$ and in the bottom panel the total current divided by the voltage $J^{\rm dc}/V$. The corresponding adiabatic coefficient has been shown in Fig. \ref{fig4}. In agreement with the analysis for small ac amplitudes, no particular symmetry 
of the total current as a function of the flux $\Phi^{\rm dc}_1$ is observed, and this behavior is not consistent with Onsager-Casimir relations. In Fig. \ref{fig9} we show results for the same setup considered in Fig. \ref{fig8}
but we now change the two fluxes   simultaneously, $\Phi^{\rm dc}_1=\Phi^{\rm dc}_2= \Phi^{\rm dc}$. We see that in this case, all the transport coefficients are even functions of $\Phi^{\rm dc}$ (recall that $G^a_{\rm dc}$ is shown in Fig. \ref{fig4}).  The behavior of the total current shown in the bottom panel of the Fig. \ref{fig9} is in the present case consistent with Onsager-Casimir relations.

In the case of two rings asymmetrically connected, $G^V_{\rm dc}$ is an even function of $\Phi^{\rm dc}_j,\; j=1,2$, but the adiabatic component does not have any defined parity
(see Fig. \ref{fig5}) and we have verified that this is also the case of the mixed component. Thus, the full dc current does not have in this case any particular symmetry
as a function of dc component of the magnetic flux.  

\section{Summary and Conclusions}
We have studied the transport properties of one and two rings threaded by magnetic fluxes with dc and ac components with and without dc bias voltage applied at
the terminals.
We have developed different theoretical strategies to solve the problem in different limits and we have defined the relevant coefficients to characterize
the transport in the limit of small bias voltages and low driving frequencies. 

We have shown that it is not possible to generate adiabatic pumping in a single ring by driving with pure ac magnetic fluxes, even for magnetic fluxes containing
several harmonics oscillating with phase-lags. This is in contrast to the behavior of a single ring threaded by a magnetic flux that changes linearly in time.
In that case, the induced constant electric field is proportional to the driving frequency and the induced dc current is, thus, proportional to the frequency even
when it corresponds to a single-parameter driving. \cite{19} In the case of  two rings  we have shown that adiabatic pumping is possible provided that the two 
magnetic fluxed have
ac components oscillating with a phase lag and also have a finite dc components, which is in agreement with previous results. \cite{doub} 

Finally, we have  analyzed the behavior of the dc currents in setups with one and two rings as the dc magnetic flux is varied with and without applied dc voltage. For the case of a single 
ring we found that the non-adiabatic pumped current does not have in general any particular symmetry as a function of the dc flux, indicating that the Onsager-Casimir relations are not valid
in this system. An arbitrary small coupling to the reservoirs in the driven ring is enough to break Onsager symmetry, except in the case where the connection is at
perfectly symmetric positions under
spacial inversion symmetry. We found a similar behavior in the case of two rings and pumping within the adiabatic regime. For the particular case of perfectly symmetric coupling to the reservoirs,
the pumped current in this setup  is even as a function of the dc magnetic flux, when the same dc flux threads the two rings, while it is an antisymmetric function under changes of the dc flux
of only one ring while keeping the constant the dc flux of the other one. The fact that in the non-adiabatic as well as in the adiabatic regimes the Onsager-Casimir relations are not expected
to be valid in general can be related to the fact that pumping is at least a second order process in the driving amplitudes as explicitly shown by Eq. (\ref{gna}) for a single ring and Eq. (\ref{2r}) for two rings. Although in the adiabatic regime the dc current  is linear in the pumping frequency it is 
 non-linear in the driving field. In this sense, the situation resembles the case of the dc driving where Onsager symmetry
is broken beyond linear response in the applied voltage. However in that case, the explanation of this breakdown resort to the effect of the interactions, \cite{viol-casimir-theor} while in the present case, it is a purely dynamical effect, which takes place even in a non-interacting system. When a dc bias voltage is, in addition applied, the dc current resulting as a combination of driving with the ac flux and with the
dc voltage is an even function of the dc magnetic flux only in the case where the setup is symmetrically connected, not showing particular symmetries in other configurations, in agreement with the
breakdown of the Onsager-Casimir relations in the pumped component of the current. The lack of symmetry of the dc current as a function of magnetic fluxes has already been discussed in systems pumped
by gate voltages. \cite{nonsim,mobu} In the present case, we have shown that some of those features also take place when the driving takes place in an oscillating component of the magnetic flux itself.

Regarding the possibility of experimental observation of the different regimes and mechanisms described in the present work, we notice that nanolitography techniques on GaAs/AlGaAs-heterostructures,
enable the fabrication of single as well as arrays of mesoscopic rings (see for example Refs. \onlinecite{2, 3, levelesp,levelesp2, arrays}). In particular, in Ref. \onlinecite{2}, the mesoscopic ring under study, which is threaded
by a magnetic flux,  is integrated in a small substrate of some hundreds of $\mu$m with a SQUID device, which is used to detect the persistent current induced in the ring. Thus, we find that the present experimental state of the art may
allow for the study of the setups and features analyzed  in the present work. In experiments, the rings have typical diameters ranging from some hundreds of nm to a few  $\mu$m, and have a mean level spacing $\Delta \sim 0.5 meV$. \cite{levelesp} In arrays of rings, the typical level spacing is estimated to be smaller  $\Delta \sim 0.001 meV$.  \cite{levelesp2} This corresponds to resonant frequencies $\Delta/h \sim  1 - 500 GHz$.  On the other hand, the range of frequencies within the adiabatic regime is 
determined by the typical width of the resonant peaks of the ring. This, in turn depends on the degree of coupling to the leads as well as the length of the ring, but it is significantly smaller than the typical level spacing. If we assume this width to be at the most $10^{-3} \Delta$, we conclude that  the adiabatic regime would correspond  to frequencies below a few hundreds of $MHz$. In the absence of a dc driving voltage, these frequencies cast dc currents from
$0.1$nA to  a few nA.  These currents are small but within the range  observed in experiments on Aharanov-Bohm rings (see \onlinecite{3,rus,levelesp,levelesp2,arrays}). 

To finalize, we would like to mention other interesting features of the transport properties of these setups. In particular, in the case of pumping in a single driven ring, we would like to stress the change
 from the  ''paramagnetic''-like behavior, characterized by a vanishing dc current for vanishing dc flux, to the ''diamagnetic''-like behavior, characterized by a finite current for zero dc flux , as the driving 
 frequency increases. This feature is akin to what has been experimentally observed in the behavior of the persistent currents in rings threaded by fluxes with ac components. \cite{3}

\section{Acknowledgements}
We thank M. Moskalets and S. Gasparinetti for interesting comments and suggestions. We also thank the Institute BIFI-Zaragoza for computer facilities.
This work is supported by CONICET, MINCyT through PICT-2010 and UBACYT, Argentina. 

\appendix

\section{Frozen Green's function for a single ring}
We consider a single ring threaded by a magnetic flux with an arbitrary number of harmonic components described 
by the Hamiltonian (\ref{ring}) and connected to reservoirs. The frozen Green's function 
can be expressed in terms of the T-matrix as
\be \label{gft}
\hat{G}^f(t,\omega)=\hat{g}^f(\omega)+\hat{g}^f(\omega)\hat{T}^f(t,\omega) \hat{g}^f(\omega),
\ee
being $\hat{g}^f(\omega)=\left[\omega \hat {1}-\hat{\Sigma}(\omega)\right]^{-1}$, and
\be
\hat{T}^f(t,\omega)=\hat{H}_r(t)+ \hat{H}_r(t)\hat{g}^f(\omega)\hat{H}_r(t)+ \ldots.
\ee
Substituting (\ref{ring}) it is possible to verify that the  T-matrix has the following structure
\be
\hat{T}^f(t,\omega)=\sum_{n=-\infty}^{+\infty}\hat{T}^{(n)}(\omega) e^{i n \phi(t)},
\ee
which, when substituted in (\ref{gft}), leads to a representation of $\hat{G}^f(t,\omega)$ in terms
of a power series of $e^{i\phi(t)}$.

\section{Frozen Green's function for two rings. Fluxes with small  ac amplitudes}
For small amplitudes of the ac components of the fluxes, we can consider the Hamiltonian (\ref{smallac}) to find that the frozen Green's function at the first order in $\phi_j(t)$ reads
\be\label{coef}
\hat{G}^f(t,\omega)= \hat{G}^0(\omega)+ \sum_{j=1}^2 \phi_j(t) \hat{G}^0(\omega) \hat{J}_j \hat{G}^0(\omega),
\ee
being $\hat{G}^0(\omega)^{-1}= \hat{1}\omega - \hat{H}_0-\hat{\Sigma}$, with $H_0$ defined in (\ref{h0}) while 
\be \label{cur}
\hat{J}_j=i \left( \hat{W}_j e^{i \phi_j^{\rm dc}}-\hat{W}_j^{\dagger} e^{-i \phi_j^{\rm dc}} \right),
\ee
defines a matrix associated to the persistent current operator of the ring $j$. 

%%%%%%%%%%%%%%%%%%%%%%%%%%%%%%%%%%%%%%%%%%%%%%%%%%%%%%%%%%%%%%%%%%%%%%%%
%%%%

%%%%%%%%%%%%%%%%%%%%%%%%%%%%%%%%%%%%%%%%%%%%%%%%%%%%%%%%%%%%%%%%%%%%%%%%
%%%%


\begin{thebibliography}{11}
\bibitem{1} V. Chandrasekhar, R. A. Webb, M. J. Brady, M. B. Ketchen, W. J. Gallagher, and A. Kleinsasser, Phys. Rev. Lett. 67, 3578 (1991).
\bibitem{2} D. Mailly, C. Chapelier, and A. Benoit, Phys. Rev. Lett. 70, 2020 (1992).
\bibitem{3} R. Deblock, R. Bel, B. Reulet, H. Bouchiat, and D. Mailly, Phys. Rev. Lett. 89, 206803 (2002).
\bibitem{4} Y. Gefen, Y. Imry and Ya. Azbel, Phys. Rev. Lett. 52, 129 (1984).
\bibitem{5} Y. Imry, ''Directions in Condensed Matter Physics'', edited by G. Grinstein, E. Mazenko (World Scientific, Singapore, 1986).
\bibitem{6} M. B\"uttiker, Y. Imry and R. Landauer, Phys. Lett. 96A, 365 (1983).
\bibitem{7} R. Landauer, IBM J. Res. Develop. 1, 233 (1957); R. Landauer, Phil. Mag. 21, 863 (1970).
\bibitem{8} M. B\"uttiker and R. Landauer, Phys. Rev. Lett. 54, 2049 (1985).
\bibitem{9} M. B\"uttiker, Phys. Rev. B 32, 1846 (1985). 
\bibitem{10} D. Lenstra and W. van Haeringen, Phys. Rev. Lett. 57, 1623 (1986). 
\bibitem{11} R. Landauer, Phys. Rev. B 33, 6497 (1986). 
\bibitem{12} R. Landauer, Phys. Rev. Lett. 58, 2150 (1987).
\bibitem{13} Y. Gefen and D. J. Thouless, Phys. Rev. Lett. 59, 1752 (1987).
\bibitem{14} D. Lubin Y. Gefen and I. Goldhirsch, Phys. Rev. B 41, 4441 (1990). 
\bibitem{15} G. Blatter and D. A. Browne, Phys. Rev. B 37, 3856 (1988). 
\bibitem{16} P. Ao, Phys. Rev. B 41, 3998 (1990). 
\bibitem{17} J. Avron and J. Nemirovsky, Phys. Rev. Lett. 68, 2212 (1992). 
\bibitem{18} M. T. Liu, and C. S. Chu, Phys. Rev. B 61, 7645 (2000). 
\bibitem{19} L. Arrachea, Phys. Rev. B 66, 045315 (2002); {\em ibid} {\bf 70} 155407 (2004);
\bibitem{19a} F. Foieri, L. Arrachea and M. J. S\'anchez,
Phys. Rev. Lett. {\bf 99}, 266601 (2007).
\bibitem{baruch}Y. Etzioni, B. Horovitz, and P. Le Doussal, Phys. Rev. Lett. {\bf 106} 166803 (2011).
\bibitem{curring}Cong-Hua Yan, Lian-Fu Wei
Jour. Phys. Cond. Matt. {\bf 22}, 185301 (2010).
\bibitem{marq}F. Marquardt and C. Bruder, Phys. Rev. B {\bf 65}, 125315 (2002).
\bibitem{spino} P. F\"oldi, Orsolya K\'alm\'an, and Mih\'aly G. Benedict
Phys. Rev. B {\bf 82}, 165322 (2010).
\bibitem{pumps}D.\ J.\ Thouless, Phys. Rev. B {\bf 27}, 6083, (1983); M.\ B\"uttiker, H.\ Thomas, and A.\ Pretre, Z. Phys. B: Cond. Mat. {\bf 94}, 133 (1994);
B.\ L.\ Altshuler and L.\ I.\ Glazman, Science {\bf 283}, 1864 (1999); M.\ Switkes {\it et al.}, Science {\bf 283}, 1905 (1999); P. Sharma and C. Chamon, Phys. Rev. Lett. {\bf
  87}, 096401 (2001); L. DiCarlo, C. M. Marcus, and J. S. Harris Jr, Phys. Rev.
Lett. {\bf 91}, 246804 (2003);M.\ D.\ Blumenthal, {\it et al.}, Nat. Phys.  {\bf 3}, 343 (2007); P.\ J.\ Leek, Phys. Rev. Lett. {\bf 95}, 256802 (2005); 
E.\ R.\ Mucciolo, C.\ Chamon, and C.\  M.\ Marcus, {\it ibid}. {\bf 89}, 146802 (2002); S.\ K.\ Watson {\it et al.}, Phys. Rev. Lett. {\bf 91}, 258301 (2003); F.\ Giazotto {\it et al.}, Nature Phys. {\bf 7}, 857 (2011).
 \bibitem{lilisim} L. Arrachea, Phys. Rev. B {\bf 72} 121306 (2005).
\bibitem{moska} M.Moskalets,M.B\"uttiker, Phys.Rev.B {\bf 66}, 035306 (2002).
\bibitem{lilipump} L. Arrachea, 
Phys. Rev. B {\bf 72}, 125349 (2005);
L. Arrachea, 
Phys Rev. B {\bf 75}, 035319 (2007);
L. Arrachea and M. Moskalets, 
Phys. Rev. B {\bf 74}, 245322 (2006).
 \bibitem{brouw}P. W. Brouwer, Phys. Rev. B {\bf 58}, R10135 (1998).
\bibitem{nonsim} T. A. Shutenko, I. L. Aleiner, and B. L. Altshuler, Phys. Rev. B {\bf 61}, 10 366 (2000).
I. L. Aleiner, B. L. Altshuler, and A. Kamenev, Phys. Rev. B {\bf 62}, 10 373 (2000).
M. Vavilov, V. Ambegaokar, and I. L. Aleiner, Physical Review B {\bf 63}, (2001). 
\bibitem{mobu}M.Moskalets,M.B\"uttiker, Phys. Rev. B {\bf 72}, 035324 (2005).
\bibitem{foa} L. E. Foa Torres, Phys. Rev. B {\bf 72}, 245339 (2005).
\bibitem{20} S. Flach, O. Yevtushenko and Y. Zolotaruk, Phys. Rev.
Lett. 84, 2358 (2000). 
\bibitem{21} S. Denisov, S. Flach, A. A. Ovchinnikov, O. Yevtushenko,
and Y. Zolotaryuk, Phys. Rev. E, 66, 041104 (2002). 
\bibitem{22} S. Flach, Y. Zolotaryuk, A. E. Miroshnickenko and M.
V. Fistul, Phys. Rev. Lett. 88, 184101 (2002). 
\bibitem{doub}D. Shin and J. Hong, Phys. Rev. B {\bf 70}, 073301 (2004).
\bibitem{sim} S. Gasparinetti and I. Kamleitner,  Phys. Rev. B {\bf 86}, 224510 (2012).
\bibitem{ben}G. Benenti, K. Saito, and G. Casati, Phys. Rev. Lett. {\bf 106}, 230602 (2011). 
\bibitem{viol-casimir-theor}M. B\"uttiker, Phys. Rev. Lett. {\bf 57}, 1761 (1986); D. S\'anchez and M. B\"uttiker, Phys. Rev. Lett. {\bf 93}, 106802 (2004); 
M. L. Polianski and M. B\"uttiker, Phys. Rev. Lett. {\bf 96}, 156804 (2006);
D. S\'anchez and K. Kang, Phys. Rev. Lett {\bf 100}, 036806 (2008);
A. R. Hern\'andez and C. H. Lewenkopf, Phys. Rev. Lett. {\bf 103}, 166801 (2009).
\bibitem{ang}L. Angers, E. Zakka-Bajjani, R. Deblock, S. Gu\'eron, H. Bouchiat, A. Cavanna, U. Gennser and M. Polianski, Phys. Rev. B {\bf 75}, 115309 (2007); L. Angers, A. Chepelianskii, R. Deblock, B. Reulet, and H. Bouchiat, Phys. Rev. B {\bf 76}, 075331 (2007).
\bibitem{viol-casimir-exp} A. L\"ofgren, C. A. Marlow, I. Shorubalko, R. P. Taylor, P. Omling, L. Samuelson, and H. Linke, Phys. Rev. Lett. {\bf 92}, 046803 (2004);
J. Wei, M. Shimogawa, Z. Wang, I. Radu, R. Dormaier, and D. H. Cobden, Phys. Rev. Lett. {\bf 95}, 256601 (2005);
C. A. Marlow, R. P. Taylor, M. Fairbanks, I. Shorubalko, and H. Linke, Phys. Rev. Lett. {\bf 96}, 116801 (2006); C. Ojeda-Aristizabal, M. Monteverde, R. Weil, M. Ferrier, S. Gu\'eron,
and H. Bouchiat, Phys. Rev. Lett. {\bf 104}, 186802 (2010).
\bibitem{sau}J. Sau, T. Kitagawa, and B. I. Halperin, Phys. Rev. B {\bf 85}, 155425 (2012).
\bibitem{rus}A. A. Bykov, Z. D. Kvon, L. V. Litvin, Yu. V. Nastaushev, V. G. Mansurov, V. P. Migal', and S. P. Moshchenko, Pis'ma Zh. Eksp. Teor. Fiz. 58, 7, 538-541 (1993).
\bibitem{levelesp}M. Cass\' e, Z. D. Kvon, G. M. Gusev, E. B. Olshanetskii, L. V. Litvin, A. V. Plotnikov, D. K. Maude and J. C. Portal, Phys. Rev. B {\bf 62}, 2624 (2000).
\bibitem{levelesp2} R. Deblock, Y. Noat, B. Reulet, H. Bouchiat, D. Mailly, Phys. Rev. B {\bf 65}, 075301 (2002).
\bibitem{arrays} W. Rabaud, L. Saminadayar, D. Mailly, K. Hasselbach, A. Benoit, B. Etienne, Phys. Rev. Lett. {\bf 86}, 3124 (2001)

\end{thebibliography}
\end{document}